\documentstyle[12pt]{article}
\newtheorem{lemma}{Lemma}

\def\tr{\mbox{tr}\,}

\def\bbbe{I\!\!\!\!E}
\begin{document}
\begin{center}
\null

{\large {\sc Bulgarian Academy of Sciences}\\
\bf  Institute for Nuclear Research and Nuclear Energy\\}
{\normalsize boul. Tzarigradsko shosse 72, 1784 Sofia, Bulgaria\\}
\vskip 3pt
{\scriptsize \bf Tel.: 003592--74311, \hfil Fax.: 003592--9753619, \hfil
 Telex: 24368 ISSPH BG \\}
\vskip 2pt
\hrule
\end{center}

\thispagestyle{empty}
\begin{flushright}
{\bf Preprint TH--97/13}
\end{flushright}

\begin{center}

{\Large \bf The complex
Toda chains and the simple Lie \\[5pt] algebras -- solutions and large
time asymptotics  } \bigskip

{\bf V S Gerdjikov, E G Evstatiev and R I Ivanov}

{\sl Institute for Nuclear Research and Nuclear Energy,
Bulg. Acad. of Sci.,\\ boul. Tzarigradsko shosse 72,
1784 Sofia,Bulgaria }

\end{center}

\bigskip
\begin{abstract}
The asymptotic regimes of the $N $-site complex Toda chain (CTC) with
fixed ends related to the classical series of simple Lie algebras are
classified.  It is shown that the CTC models have much richer variety of
asymptotic regimes than the real Toda chain (RTC). Besides asymptotically
free propagation (the only possible regime for the RTC), CTC allow bound
state regimes, various intermediate regimes when one (or several) group(s)
of particles form bound state(s),  singular and degenerate solutions.
These results can be used e.g., in describing the soliton interactions of
the nonlinear Schr\"odinger equation. Explicit expressions for the
solutions in terms of minimal sets of scattering data are proposed for all
classical series ${\bf B}_r $-- ${\bf D}_r $.

\end{abstract}
\vfill
\centerline{Revized version -- February 1998}

\section{Introduction}
The Toda chain model
\cite{Man,Fla,Moser,LeSav1,LeSav,Toda,Toda1,Bogoy1,OlshPer,%
OlshPer1,MihOlPer1,Mih1}:
\begin{equation}\label{eq:CTC}
{ d^2 q_k \over  d t^2 } = \exp{(q_{k+1} - q_k)} -
\exp{(q_{k} - q_{k-1})},
\end{equation}
is one of the paradigms of integrable nonlinear chains and lattices.  It
has been thoroughly studied for a number of initial and boundary
conditions, such as:
\begin{itemize}
\item  fixed ends boundary conditions, i.e. $q_0= - q_{N+1} =\infty $;
this will be the case we are interested in;

\item infinite chain $-\infty < k < \infty  $ with $\lim_{k\to - \infty }
q_k =0 $ and $\lim_{k\to \infty } q_k = \mbox{const}$, or equivalently,
$\lim_{k\to \pm \infty } (q_{k+1} - q_k) =0 $;

\item quasi-periodic boundary conditions $q_{k+N} = q_{k} + c $, where
$c=\mbox{const} $.
\end{itemize}

This model appeared first in describing the oscillations of a
one-dimensional crystalline lattice \cite{Toda1}. Since then many other
applications became known, see e.g. \cite{Toda}.

The model (\ref{eq:CTC}) is directly related to the algebra $sl(N) $,
where $N $ is the number of sites of the chain. Most of the references
cited above are devoted to the case when $q_k(t) $ are real-valued
functions. That is why for definiteness we will call this model real Toda
chain (RTC). Other generalizations of the RTC are related to:  a)~simple
Lie algebras; b)~affine (or Kac-Moody) algebras; c)~2-dimensional
generalizations.

Another possibility for generalizing the RTC, which as far as we know
has not been investigated, is the complex Toda chain (CTC). We see two
main reasons for this:

(i) the generic solutions of the CTC are readily obtained from the ones of
RTC by simply making all parameters complex. In fact, technically solving
the RTC requires  additional efforts in ensuring that the scattering data
of the Lax matrix $L $ are real-valued.

(ii) the CTC has not been known to have physical applications.

Recently however, it was discovered \cite{GKUE,GUED,Arnold} that the CTC
describes the $N $-soliton train interaction of the nonlinear
Schr\"odinger (NLS) equation in the adiabatic approximation. The variables
$q_k(t) $ in  (\ref{eq:CTC}) are related to the solitons parameters by
\begin{eqnarray}\label{eq:q-k}
q_k = - 2\nu_0  \xi_k   +   i \left( 2\mu_0 \xi_k - \delta_k \right) +
\mbox{const}
\end{eqnarray}
where $\xi_k $ and $\delta _k $ characterize the center-of-mass position
and the phase of the $k $-th soliton in the chain; $\nu _0 $ and $\mu _0 $
are the average amplitude and velocity of the soliton train.  Such soliton
trains and their asymptotic behavior appear to be important for the needs
of soliton based fiber optics communications.

In what follows we will view the CTC as a model of ``complex'' particles
characterized by a coordinate and phase which, in analogy with
(\ref{eq:q-k}),  are related to $\mbox{Re\,}q_k(t) $ and
$\mbox{Im\,}q_k(t) $ respectively.

Another reason for the present paper is in the fact, that along with the
similarities between the solutions of CTC and RTC, there are also
important qualitative differences between the asymptotic properties of
these solutions.

The purpose of the present paper is to derive analytically the  large-
time asymptotics of the solutions to the CTC models:
\begin{equation}\label{eq:gCTC}
{ d^2 q\over  d t^2 } \equiv \sum_{k=1}^{r}
{ d^2 q_k  \over  d t^2 } H_k
= \sum_{\alpha \in \pi_g}^{} H_\alpha   e^{-(\vec{q},\alpha) },
\end{equation}
related to the classical series of simple Lie algebras extending the
results of \cite{Goodman,OPRS}.  Here $\pi_g \equiv \{ \alpha _1, \dots,
\alpha _r\} $ is the set of simple roots of the algebra $\sl g $, $q(t) $
is a complex-valued function of $t $ taking values in the Cartan
subalgebra $\sl h $ of $\sl g $; $H_k $ form a basis in $\sl h $ dual to
the orthonormal basis $\{ e_k\}_{k=1}^{r} $ in the root space $\bbbe^r $
and $r $ is the rank of $\sl g $; $H_\alpha = \sum_{k=1}^{r} (\alpha ,e_k)
H_k$. For more details about the structure of the simple Lie algebras, see
e.g. \cite{Bour}. Equation (\ref{eq:CTC}) is a particular case of
(\ref{eq:gCTC}) for the ${\bf A}_r $ series, i.e., for ${\sl g}\simeq
sl(r+1) $.  These results, and especially the ones for the series ${\bf
A}_r $, can be used as a tool for deriving the asymptotic behavior of the
$N $-soliton trains from the initial set of soliton parameters
\cite{GKUE,GUED,Arnold}.

We also specify the minimal sets of scattering data ${\cal  T}_{g} $ for
$L $ which determine uniquely both $L $ and the solutions of
(\ref{eq:gCTC}) and obtain explicit expressions for the solutions of
(\ref{eq:gCTC}) in terms of ${\cal T}_g $.

\section{Comparison between RTC and CTC}

Since the paper of Moser \cite{Moser} on the finite nonperiodic real Toda
lattice, there have been proposed many methods for solving (\ref{eq:CTC})
for the various choices of the initial and boundary conditions, see
\cite{Man,Fla,Moser,LeSav1,LeSav,Toda,Toda1,Bogoy1,OlshPer,%
OlshPer1,MihOlPer1} and the numerous references therein.  The RTC model
was also extended to indefinite metric spaces \cite{Kod} and was shown to
possess singular solutions ``blowing up'' for finite values of $t $. These
models can be viewed as special cases of the CTC in which part of the $a_k
$'s defined below are real while the rest are purely imaginary.

As we already mentioned, a number of properties of the CTC are obtained
from the corresponding ones of RTC trivially by assuming the corresponding
variables complex. We list the four most important of them below:

S1)~The Lax representation:  There are several equivalent formulations of
the Lax representation for (\ref{eq:gCTC}). Below we will use the
``symmetric'' one:
\begin{equation}\label{eq:g-laxa}
L(t)=\sum_{k=1}^{r} \left( b_k H_k + a_k(E_{\alpha_k }+
E_{-\alpha_k }) \right),
\end{equation}
\begin{equation}\label{eq:g-laxb}
M(t)= {1\over 2}\sum_{k=1}^r a_k(E_{\alpha_k } - E_{-\alpha_k }),
\end{equation}
where $a_k={1 \over 2}\exp{(-(q,\alpha_k)/2)} $ and $b_k={1\over 2} { d
q_k/ d t }$; for ${\sl g}\simeq sl(N) $ we have $a_k={1 \over
2}\exp{((q_{k+1}-q_k)/2)} $.  It is well known that to each root $\alpha
\in \Delta _g\subset \bbbe^r $ one can put into correspondence the element
$H_\alpha \in {\sl h} $.  Analogously, to $q(t) = \mbox{Re\,} q(t)+ i
\,\mbox{Im\,} q(t) $ there corresponds the vector $\vec{q}(t) =
\mbox{Re\,} \vec{q}(t)+  i\, \mbox{Im\,} \vec{q}(t) $, whose real and
imaginary parts are vectors in $\bbbe^r $.

S2)~The integrals of motion in involution are provided by the eigenvalues,
$\zeta_k $, of $L $.

S3)~The  solutions of both the CTC and the RTC are determined by the
scattering data for $L_0\equiv L(0) $. When the spectrum of $L_0 $ is
nondegenerate, i.e.  $\zeta _k \neq \zeta _j $ for $k\neq j $, then this
scattering data consists of
\begin{equation}\label{eq:Tau}
{\cal  T}\equiv \{ \zeta _1, \dots , \zeta _N, r_1, \dots, r_N\}.
\end{equation}
where $r_k $ are the first components of the corresponding eigenvectors
$r_{k} = V_{1k} $ of $L_0 $ in the typical representation $R(\omega _1) $
of $\sl g $, $N=\mbox{dim\,}R(\omega_1) $:
\begin{equation}\label{eq:eigve}
L_0 V= V Z, \qquad Z=\mbox{diag\,}(\zeta _1, \dots, \zeta_N).
\end{equation}
They are determined (up to an overall sign) by the normalization
conditions:
\begin{equation}\label{eq:norm}
\sum_{k=1}^{N} \left( V_{sk}\right)^2 = (v^{(s)}, v^{(s)}) = 1,
\qquad s=1,\dots, N,
\end{equation}
see \cite{Moser,LeSav,Toda}; then $V^T=V^{-1} $.

S4)~Lastly, the eigenvalues of $L_0$ uniquely determine the asymptotic
behavior of the solutions; these eigenvalues can be calculated directly
from the initial conditions. We will extensively use this fact for the
description of the different types of asymptotic behavior.

However, there are important differences between the RTC and CTC,
especially the asymptotic behavior of their solutions. Indeed, for the
RTC, one has  \cite{Moser,Toda} that, both the eigenvalues, $\zeta _k $,
and the coefficients, $r_k $, are always real-valued. Moreover, one can
prove that $\zeta _k \neq \zeta _j $ for $k\neq j $, i.e. no two
eigenvalues can be exactly the same.  As a direct consequence of this, it
follows that the only possible asymptotic behavior in the RTC is an
asymptotically separating, free motion of the particles (solitons).

This situation is different for the CTC. Now the eigenvalues $\zeta_k=
\kappa _k +  i\eta_k$, as well as the coefficients $r_k$ become complex.
Furthermore, the argument of Moser can not be applied so one can  have
multiple eigenvalues.  The collection of eigenvalues, $\zeta _k$, still
determines the asymptotic behavior of the solutions. In particular, it is
$\kappa _k $ that determines the asymptotic velocity of the $k $-th
particle (soliton).  For simplicity, we assume $\zeta _k\neq \zeta_j$ for
$k\neq j $.  However, this condition does not necessarily mean that
$\kappa _k \neq \kappa _j $.  We also assume that the $\kappa _k$'s are
ordered as:
\begin{equation}\label{eq:kapp}
\kappa _1 \leq \kappa_2 \leq \dots \leq \kappa _N .
\end{equation}
Once this is done, then for the corresponding set of $N$ particles (train
of $N$ solitons), there are three possible general configurations:

D1) $\kappa _k\neq \kappa _j $ for $k\neq j $.  Since the asymptotic
velocities are all different, one has the asymptotically separating, free
particles (solitons).

D2)~$\kappa _1 =  \kappa _2 =\dots = \kappa _N $. In this case, all $N $
particles (solitons) will move with the same mean asymptotic velocity, and
therefore will form a ``bound state".  The key question now will be the
nature of the internal motions in such a bound state.

D3)~One may have also a variety of intermediate situations when only one
group (or several groups) of particles move with the same mean asymptotic
velocity; then they would form one (or several) bound state(s) and the
rest of the particles will have free asymptotic motion.

Obviously the cases D2) and D3) have no analogues in the RTC and
physically are qualitatively different from D1).  The same is also true
for the special degenerate cases, where two or more of the $\zeta _k $'s
may become equal and for the singular solutions.  These cases will be
considered briefly below.

\section{Solutions of the CTC}

The solutions for the CTC can be obtained formally from the well known
ones for RTC by inserting the corresponding complex parameters.  We remind
first the solution of the RTC for ${\sl g}\simeq sl(N) $, see
\cite{Moser,Toda,And} and the references therein.  Here we fix up the mass
center at the origin by
\begin{equation}\label{eq:massc}
\sum_{k=1}^{N} q_k(t)=0.
\end{equation}
The velocity of the center of mass is given by $\tr L_0 = \sum_{k=1}^{N}
\zeta _k =0 $, due to $L\in sl(N) $.  Then the solution has the form:
\begin{equation}\label{eq:qaa}
q_k(t)=q_1(0) + \ln {A_{k}\over A_{k-1}}
\end{equation}
where $A_0\equiv 1 $,
\begin{equation}\label{eq:A1}
A_1(t)=\sum_{k=1}^{N}r_{k}^{2} e^{-2\zeta_k t}
\end{equation}
\begin{eqnarray}\label{eq:Ak}
A_{k}(t)= \sum_{1\leq l_{1}<l_{2}<\ldots<l_{k}\leq N}
(r_{l_1}r_{l_2}\ldots r_{l_k})^{2} W^{2}(l_1,l_2,\ldots ,l_k)
 e^{-2(\zeta _{l_1} + \dots + \zeta _{l_k})t}
\end{eqnarray}
 and
\begin{equation}\label{eq:A_NConst}
 A_N =  W^2(1,2,\dots,N) \prod_{k=1}^{N} r_k^2=\exp{(-Nq_1(0))}.
\end{equation}
By $W(l_1,\dots, l_k) $ we denote the Wandermonde determinant:
\begin{equation}\label{eq:Wan}
 W(l_1,\dots , l_k) = \prod_{\begin{array}{c} \scriptsize s>p \\
\scriptsize s,p \in \{ l_1,\dots , l_k\}\end{array}}
(2\zeta _s - 2\zeta _p).
\end{equation}

The solutions of (\ref{eq:CTC}) for real-valued $q(t) $ are well known in
several different formulations. We remind here the formula, which
effectively is contained in \cite{OlshPer}:
\begin{equation}\label{eq:q-ome}
(\vec{q}(t),\omega _k) - (\vec{q}(0),\omega _k) =
\ln \langle \omega_k | \exp{(-2L_0 t)} |\omega_k \rangle ,
\end{equation}
where $\omega _k $ are the fundamental weights of $\sl g $. The fact that
the large time asymptotics of $q_k(t) $ for the ${\sl g}\simeq sl(N) $ RTC
have the form:
\begin{equation}\label{eq:asy-sl}
\lim_{t\to\pm\infty } (q_k(t) - v _k^\pm t) = \beta _{k}^{\pm}
\end{equation}
where the asymptotic velocities $v_k^+=- 2\zeta _k $ and $v_k^-=- 2\zeta
_{N+1-k} $ have been derived by Moser \cite{Moser}. He also evaluated the
differences $\beta _{k}^{+} -\beta _{N-k+1}^{-} $ which characterizes the
interaction in the RTC model.

The minimal set of scattering data for ${\sl g}\simeq sl(N) $  is obtained
from (\ref{eq:Tau}) by imposing on ${\cal  T} $ the restrictions
$\sum_{k=1}^{N}\zeta _k =0 $ and
\begin{equation}\label{eq:norm1}
\sum_{k=1}^{N} r_k^2 =1.
\end{equation}
Note that $\exp{(-q_1(0))} $ is expressed through ${\cal  T} $ by
(\ref{eq:A_NConst}).

For the RTC related to the other classical series of Lie algebras it
is known \cite{Goodman,OPRS} that
\begin{equation}\label{eq:asymp}
\lim_{t\to\pm\infty } (q(t) - v^{\pm} t) = \beta ^{\pm},
\end{equation}
$v^{\pm}\in {\sl h}$, $\beta ^{\pm} \in {\sl h}$ and $v^{+}=w_0(v^{-})$,
where $w_0$ is the element of the Weyl group, which maps
the highest weight of each irreducible representation of ${\sl g}$
into the lowest weight. The action of $w_0$ on the simple roots is
well known \cite{Bour}:
\begin{equation}\label{eq:w0}
w_0(\alpha _k)=-\alpha _{\tilde{k}},
\end{equation}
$\tilde{k}=r-k+1$ for ${\bf A}_r$, $\tilde{k}=k$,  $k=1,\ldots,r$ for
${\bf B}_r$, ${\bf C}_r$ and, when $r$ is even, also for ${\bf D}_r$. When
${\sl g}\simeq {\bf D}_r $ and $r$ is odd $\tilde{k}=k$ for $k\leq r-2$,
and $w_0(\alpha _{r-1})=-\alpha _r $,  $w_0(\alpha _r)=-\alpha _{r-1} $.

What we will do below is to: i)~specify how minimal sets of scattering
data ${\cal  T}_g $ can be extracted from (\ref{eq:Tau}); ii) find
explicit expressions for $\beta _k^\pm \equiv (\beta ^{\pm}, e_k) $ for
each of the classical Lie algebras in terms of ${\cal  T}_g $.

Like in the $sl(N) $-case $\zeta _k $ and $r_k $ are the eigenvalues and
the first components of the eigenvectors of $L_0 $ in the typical
representation, namely:
\begin{equation}\label{eq:geigve}
L_0 V= V \sum_{k=1}^{r}\zeta _k H_{k},
\end{equation}
The requirement that $L_0 $ (and as consequence, $L $) belongs to one of
the algebras in the ${\bf B}_r $ or ${\bf C}_r $ series imposes on $q_k $
the following natural restrictions:
\begin{equation}\label{eq:inv-a}
q_k = -q_{N-k+1}
\end{equation}
which leads to
\begin{equation}\label{eq:inv-b}
a_k = a_{N-k}
\end{equation}
\begin{equation}\label{eq:inv-c}
b_k = -b_{N+1-k}.
\end{equation}

Thus the solutions for ${\sl g}\simeq {\bf B}_r $ and ${\bf C}_r $ can
formally be obtained from the ones for $sl(N) $
(\ref{eq:qaa})--(\ref{eq:A_NConst}) by imposing on them the involutions
(\ref{eq:inv-a})--(\ref{eq:inv-c}). So we have to find out what are the
restrictions on ${\cal T} $ imposed by (\ref{eq:inv-a})--(\ref{eq:inv-c});
this will provide us with the corresponding minimal set of scattering data
${\cal T}_{g}$, which must obviously contain only $2r $ parameters.  It is
easy to find that $\zeta _{\bar{k}} = -\zeta_k $, so only $r $ of them are
independent.  It is not so trivial to derive the corresponding relations
which reduce the number of the coefficients $r_k $.  Our analysis shows
that:
\begin{equation}\label{eq:inv-r}
r_k r_{\bar{k}}=\exp{(-q_1(0))} w_k, \qquad k=1,\dots, r
\end{equation}
where $\bar{k}=N+1-k $ and $w_k $ is expressed in terms of $\zeta _k $.
Below we provide the explicit formulas for $w_k $ for  each of the
classical series ${\bf B}_r $, ${\bf C}_r $ and ${\bf D}_r $.

{\bf ${\bf B}_r $-series: $N=2r+1 $.} Note that in this case $\zeta
_{r+1}=0 $, and in addition to (\ref{eq:inv-r}):
\begin{equation}\label{eq:r0-b}
r_{r+1}^2 = \exp{(-q_1(0))} {1\over{2^{2r}(\zeta_1 \zeta_2 \dots
\zeta_r)^2}}
\end{equation}
and the expression for $w_k $ is provided by:
\begin{equation}\label{eq:w-k-b}
w_k = {1  \over  8\zeta _k^{2}} \prod_{s=1}^{k-1} {1  \over  4\zeta_s^2
- 4\zeta_k^2} \prod_{s=k+1}^{r} {1  \over 4\zeta _k^2 - 4\zeta _s^2},
\end{equation}
Inserting (\ref{eq:inv-r})--(\ref{eq:w-k-b}) into (\ref{eq:norm1}) we
obtain a quadratic equation for $\exp{(-q_1(0))} $, so it can be expressed
in terms of ${\cal T}_{g}$.

{\bf ${\bf C}_r $-series: $N=2r $.} Here
\begin{equation}\label{eq:w-k-c}
w_k = -{1  \over  4\zeta _k} \prod_{s=1}^{k-1} {1  \over  4\zeta _s^2
- 4\zeta_k^2} \prod_{s=k+1}^{r} {1  \over 4\zeta _k^2 - 4\zeta _s^2},
\end{equation}
Like for the ${\bf B}_r $-series $\exp{(-q_1(0))} $ is determined from
(\ref{eq:norm1}).

{\bf ${\bf D}_r $-series: $N=2r $.} Here
\begin{equation}\label{eq:w-k-d}
w_k = \prod_{s=1}^{k-1} {1  \over 4\zeta _s^2
- 4\zeta_k^2} \prod_{s=k+1}^{r} {1  \over 4\zeta _k^2 - 4\zeta _s^2},
\end{equation}
Again $\exp{(-q_1(0))} $ is determined from (\ref{eq:norm1}) and
(\ref{eq:inv-r}).  The derivation of the solution for this series requires
some additional efforts. The main problem here is to find explicit
parametrization for the right hand sides of (\ref{eq:q-ome}) for $k=r- 1 $
and $r $ in terms of $r_k $, which characterize the matrix $V $ in
(\ref{eq:geigve}) in the typical representation. Skipping the details we
just present the result, namely that $q_k $ with $k=1,\dots,r-1 $ are
again given by (\ref{eq:qaa})--(\ref{eq:Ak}) where $\zeta _k=- \zeta
_{\bar{k}} $ and $r_k $ are restricted by (\ref{eq:inv-r}),
(\ref{eq:w-k-d}). For $k=r $ the solution for $q_r(t) $ is provided by
(\ref{eq:qaa}) with
\begin{eqnarray}\label{eq:Dr}
A_{r}(t) = \sum_{1\leq l_{1}<l_{2}<\ldots<l_{r}\leq N}
(r_{l_1}r_{l_2}\ldots r_{l_r})^{2}  W^{2}(l_1,l_2,\ldots ,l_r)
f^2_{l_1,\dots,l_r}  e^{-2(\zeta _{l_1} + \dots + \zeta _{l_r})t}
\end{eqnarray}
and
\begin{equation}\label{eq:flr}
f_{l_1,\dots, l_r} ={1\over 2}\left( 1 +   {\zeta _1 \zeta _2 \dots \zeta
_r \over \zeta _{l_1} \zeta _{l_2} \dots \zeta _{l_r}} \right).
\end{equation}
The proof of these formulas is based on detailed analysis of the
properties of the fundamental representations $R(\omega _k) $ of ${\sl g}
$ and of their tensor products.

We remind that these formulas are valid both for RTC and CTC. In the
latter case one should be careful to avoid the subset of singular cases,
when one or more of the functions $A_k(t) $ may develop zeroes for finite
values of $t $, see also the discussion in Section~V below.

\section{Large time asymptotics }
Let us now express large time asymptotics of $q_k(t) $ in terms of the
minimal set of scattering data ${\cal  T}_g$ and analyze the different
types of asymptotic regimes.

As we mentioned above, we view the CTC as a model,  describing nontrivial
scattering of $N $ ``complex particles'' so that $\mbox{Re\,}q_k(t) $ and
$\mbox{Im\,}q_k(t) $ correspond to their coordinates and ``phases''.

{\bf D1)  Asymptotically free states  $\kappa _k \neq \kappa _j $ for
$k\neq j $.} It is specified by imposing on $ \zeta _{k} = \kappa _k +
 i \eta_k $, $k=1,\dots,N $ the so-called sorting condition:
\begin{equation}\label{eq:sort}
\kappa _1 < \kappa _2 < \dots < \kappa _N.
\end{equation}

Now we have to express $\beta _k^\pm $ in terms of ${\cal  T}_g $.
Skipping the details we list the results  for the classical series of
simple Lie algebras ${\bf A}_r $ -- ${\bf D}_r $.

{\bf ${\bf A}_r $-series: $N=r+1 $.} The corresponding ${\cal  T}_g $ is
formed from $\{ \zeta _k, r_k \}_{k=1}^{N} $ by imposing $\sum_{k=1}^{N}
\zeta _k =0 $ and the normalization condition (\ref{eq:norm1}).  Then
\begin{equation}\label{eq:A-ra}
\beta _k^+ = q_1(0) + \ln r_{k}^2 + \ln \prod_{s=1}^{k-1} (2\zeta _s -
2\zeta _k)^2
\end{equation}
\begin{equation}\label{eq:A-rb}
\beta _k^- =  q_1(0) + \ln r_{\bar{k}}^2 + \ln \prod_{s=\bar{k}+1}^{N}
(2\zeta _s - 2\zeta _{\bar{k}})^2,
\end{equation}
where $\bar{k}=N+1-k $.

Now it is easy to calculate the shift of the relative position which is
the effect of the particle interaction.  Naturally these shifts are also
complex-valued. If $q_k $ correspond to NLS solitons then $\beta _k^+
-\beta _k^- $ will describe the shifts of both the relative positions and
phases of the solitons, see formula (\ref{eq:q-k}).  Note that in the
class of regular solutions the particles do not collide, i.e. their
trajectories do not intersect. Since we have ordered the particles by
their velocities assuming $\kappa _1 < \kappa _2 < \dots < \kappa _N $, so
for $t\to \infty $ the $k $-th particle will move with velocity $-2\kappa
_k $. For $t\to -\infty  $ however we find that due to the interaction now
the $k $-th particle moves with velocity $-2\kappa _{N-k+1}=-2\kappa
_{\bar{k}} $.  This is the so-called ``sorting property'' characteristic
for the RTC.  If we identify the $k $-th particle (soliton) by its
velocity then its shift of position will be given by the real part of the
relation:
\begin{eqnarray}\label{eq:6.3}
&& \beta_{\bar{k}}^{-} - \beta_k^{+} = \sum_{j\neq k}
\epsilon _{jk} \ln(2\zeta_j - 2\zeta_k)^2,
\end{eqnarray}
where $\epsilon _{jk} =1$ for $j>k $ and $\epsilon _{jk} =-1$ for $j<k $.
The imaginary part of (\ref{eq:6.3}) will provide the shift in the phase of
the corresponding particle (soliton). Equation (\ref{eq:6.3}) is a natural
generalization of the corresponding result of Moser \cite{Moser} for RTC.

The asymptotics for the ${\bf B}_r $, ${\bf C}_r $ and ${\bf D}_r $ series
have the form (\ref{eq:asymp}) :
\begin{equation}\label{eq:asymp2}
\lim_{t\to\pm\infty } (q_k(t) \pm 2 \zeta _k t) = \beta _{k}^{\pm}.
\end{equation}
with
\begin{equation}\label{eq:Bbk-a}
\beta _k^+ = q_1(0) + \ln r_k^2 + \ln \prod_{s=1}^{k-1} (2\zeta _s -
2\zeta _k)^2
\end{equation}
\begin{equation}\label{eq:Bbk-b}
\beta _k^- = -q_1(0) + \ln {w_k^2 \over r_k^2} + \ln
\prod_{s=1}^{k-1} (2\zeta _s - 2\zeta _k)^2
\end{equation}
for $k=1,\dots,r $.

The only exception is for the series ${\bf D}_r $ in the case of odd $r$,
$k=r $ and $t\to -\infty $ which reads:
\begin{equation}\label{eq:asympr}
\lim_{t\to-\infty } (q_r(t) + 2 \zeta _r t) = \beta _{r}^{-}
\end{equation}
where
\begin{equation}\label{eq:Dbk-c}
\beta _r^- = q_1(0) + \ln{ r_r^2 } + \ln \prod_{s=1}^{r-1} (2\zeta _r -
2\zeta _{\bar{s}})^2.
\end{equation}

{\bf D2) Bound states: $\kappa _1=\kappa _2=\dots =\kappa _N=0 $.} Now
all $N $ particles will move with the same mean asymptotic velocity for
both $t\to \infty  $ and $t\to -\infty  $; by Galilean transformation this
velocity can always be made 0. The individual velocities of the particles
oscillate around the common mean value. In other words we find that all $N
$ particles generically do not separate but form a bound state with a
large number of degrees of freedom.  The explicit solutions for $q_k $ now
do not simplify even for $ t\to \pm \infty  $.  Nevertheless two features
are worth noting.

The solutions will be periodic functions of $t $ provided the ratios
$(\eta_k -\eta_m)/(\eta_k -\eta_j) $ are rational numbers for all $k $, $m
$ and $j $. In some cases they can also become singular, see the next
section.

Important for possible physical applications is the so-called
quasi-equidistant regime in which the distances between the neighboring
particles, i.e. $\mbox{Re\,}(q_{k+1} (t) -q_k(t)) $ oscillate with very
small amplitude; with rather good accuracy they could be considered
constant.  The fact that such regimes are possible and describe adequately
the behavior of certain NLS soliton trains is shown in \cite{solv-int}.

{\bf D3) Mixed regimes.} As we mentioned above there are a number of
intermediate cases. Here we start with the case when $m+1 $ out of
the $N$ particles form a bound state, i.e.:
\begin{equation}\label{eq:iii}
\kappa _1<\dots <\kappa _k = \dots =\kappa _{k+m} < \kappa
_{k+m+1} <\dots <\kappa _{N}
\end{equation}
and $\eta_i \neq \eta_j $ for $i\neq j \in \{ k, k+1, \dots k+m\} $.
Skipping the details we present the results for the case ${\sl g}\simeq
sl(N) $ and $m=1 $, i.e., only two of the particles form a bound state.
Particles with numbers different from $k$, $k+1$ are free, and for $k$-th
and $k+1$-st have
\begin{equation}\label{eq:q-kpa}
q_{k+a}(t) =q_1(0)+ u_{k}t+ \beta _{k+a}^{+}(t) +  {\cal O}( e^{-\nu t}),
\end{equation}
for $t\to \infty  $ and
\begin{equation}\label{eq:q-kma}
q_{N-k+a}(t) =q_1(0)+ u_{k}t+ \beta _{k+a}^{-}(t) +  {\cal O}( e^{\nu t}),
\end{equation}
for $t\to -\infty  $. Here $a=0,1 $, $u_k = -(\zeta _{k}+\zeta_{k+1}) $,
$$\nu = \min (\kappa  _{k}- \kappa _{k-1}, \kappa _{k+2}-\kappa_{k+1}), $$
and
\begin{equation}\label{eq:beta-k}
\beta _{k+a}^{\pm}= \tilde{\beta }_{k}^{\pm} + (-1)^a
B_{k}^{\pm}(t) + a \ln (2\zeta _k - 2\zeta _{k+1})^2,
\end{equation}
\begin{equation}\label{eq:B-k}
B_{k}^{\pm}(t) = \ln\left( 2\cos ((\eta_k - \eta_{k+1})t -  i
\gamma_{k}^{\pm} ) \right),
\end{equation}
\begin{equation}\label{eq:beta-kt}
\tilde{\beta }_{k}^{\pm} = \ln \left( r_k r_{k+1}
\mathop{\prod_{s,k}}\nolimits^\pm (2\zeta _s - 2\zeta _k) (2\zeta _s -
2\zeta _{k+1}) \right)
\end{equation}
\begin{equation}\label{eq:13.3m}
\gamma^{\pm}_k = \ln \left( {r_{k+1}\over r_k}\mathop{\prod_{s,k}}
\nolimits^\pm { \zeta_s - \zeta_{k+1} \over \zeta_s - \zeta_{k}} \right),
\end{equation}
where  $\mathop{\prod_{s,k}}\nolimits^+ \equiv \prod_{s=1}^{k-1} $ and
$\mathop{\prod_{s,k}}\nolimits^- \equiv \prod_{s=k+2}^{N} $.

\section{On the singular and degenerate solutions of CTC}

It is only natural that some of the CTC solutions do not enjoy all the
good properties of the RTC ones. We mentioned already two of them:

P1)~The elements of ${\cal  T}_g $ for the RTC are real-valued;

P2)~The eigenvalues $\zeta _k $ are pairwise different, i.e. $\zeta
_k\neq\zeta _j $ for $k\neq j $.

An immediate consequence of P1) and P2) is the fact that the solutions
$q_k(t) $ of the RTC are regular functions for all finite values of
$t $.

Generalizing to CTC we loose both properties P1) and P2). As a result,
besides the regular solutions, we obtain also singular and degenerate
solutions. Obviously, all solutions leading to the D1 regime are regular.
Even if we assume that property P2) holds, we may still have singular
solutions.

Indeed, from equations (\ref{eq:B-k}) -- (\ref{eq:13.3m}) we see that in
the ``oscillating part'' of the motion $B_k^\pm(t) $ is periodic.  If in
addition the parameters in ${\cal  T}_g $ are such that $\mbox{Re\,}\gamma
_{k}^{\pm}=0 $ in (\ref{eq:13.3m}) then $B_k(t) $ will develop
singularities for finite values of $t $. The same holds true for the
functions $q_k(t) $: there exist submanifolds of ${\cal  T}_g $ for which
$q_k(t) $ become singular for finite values of $t $. This fact has been
noted in \cite{Kod} for RTC on spaces with indefinite metric and (or) with
purely imaginary interaction constant.  Such RTC can be viewed as
particular case of the CTC when one or several of the functions $a_k(t) $
are purely imaginary, while the others remain real.

Let us now examine the degeneration, i. e. the case when the property P2)
is violated and some of the eigenvalues of $L_0$ become equal.  In
\cite{Moser,Trub} it is proved that the spectrum of $L_0$ is simple for
real $sl(N)$-Toda chain.  We state the following generalization:

\begin{lemma}\label{lem:1}
Let us consider $L_0$ for the complex Toda models related
to the classical series of simple Lie algebras $A_r$, $B_r$, $C_r$,
$D_r$.
If $L_0$ does not contain Jordan cells, then its spectrum is simple.
\end{lemma}

Therefore the degeneration can take place only for CTC models and only
provided in the diagonalization of $L_0 $ Jordan cells are present.  Let
us have ${\sl g}\simeq sl(N) $ and let $L_0$ has a $2\times 2$ Jordan
cell, $\zeta _1=\zeta _2= \zeta $.  Then $ L_0 $ has an eigenvector $v
^{(1)}$ and an adjoint eigenvector $v ^{(2)}$:
\begin{equation}\label{eq:dega}
L_0v^{(1)}=\zeta v^{(1)} ,
\end{equation}
\begin{equation}\label{eq:degb}
L_0v^{(2)}=\zeta v^{(2)} + v^{(1)},
\end{equation}
where $v ^{(2)}$ can be expressed as linear combination of $v^{(1)}(\zeta
) $ and its first derivative with respect to $\zeta  $. The corresponding
$A_k(t) $ besides the standard exponential terms will contain also terms
of the form $t e^{-2\zeta t} $.  More generally, if the degeneracy is of
higher order, i.e., $\zeta _1=\dots=\zeta _m $ then we will need linear
combinations of $v^{(1)}(\zeta ) $ and its derivatives with respect to
$\zeta  $ of order $1 $, \dots, $m-1 $ and $A_k $ will contain terms of
the form $t^p  e^{-2n\zeta t} $ with $p=1,\dots,m-1 $ and
$n=1,\dots,\min(k,m) $, see also \cite{FarMin}.

In particular, if we have complete degeneracy (i.e., all $\zeta_k$ are
equal and equal to zero) the solution of the $sl(N) $-CTC  is expressed
through $A_k$, which are polynomials of degree  $k(N-k)$. Their
coefficients depend on $N-1 $ constants $f_k $, $k=1 ,\ldots, N-1 $.  For
example, for $N=3$ and $\zeta_1=\zeta _2=\zeta _3=0 $ we get:
\begin{equation}\label{eq:q12}
A_{1}(t)= -{1\over 2}t^{2}+f _{1}t+f _{2},
\end{equation}
\begin{equation}\label{eq:q22}
A_{2}(t)= -{1\over 2}t^{2}+f _{1}t -f _{1}^{2}-f _{2}.
\end{equation}
and $A_3=1 $. Obviously, these solutions will be regular if $A_k $ have
complex roots and will develop singularities if one (or more) of their
roots are real.

\section{Conclusions}
We have shown that the $N $-site CTC with fixed ends has richer variety of
asymptotical regimes than the RTC.

In particular, we showed that the CTC allows solutions in which the
``complex'' particles form regular bound states. This could be important
for the applications to soliton interactions, see e.g. \cite{solv-int}
where this is done for $N=3$.  Then the problem to determine the initial
soliton parameters (i.e., the values of $a_k $ and $b_k $) is reduced to
the analysis of the characteristic equation for $L_0 $:
\begin{equation}\label{eq:ch-L}
\det (L_0 - \zeta ) = \sum_{k=0}^{N} p_k \zeta ^k
\end{equation}
$p_N=1 $, $p_{N-1} =0 $ and to the requirement that the eigenvalues of
$L_0 $ be purely imaginary. This is an algebraic problem which often can
be solved analytically. It allows one to determine the set of initial
soliton parameters in such a way, that the solitons will not only form a
stable bound state, but also will propagate quasi- equidistantly.  This
type of propagation is of importance for soliton based fiber optics
communications \cite{solv-int}.

We have studied also the asymptotic regimes for the CTC related to the
other simple Lie algebras. We proposed explicit solutions to these models
in terms of the minimal sets of scattering data ${\cal  T}_g $. The
degenerate and singular solutions of the CTC which have no counterparts in
RTC are also briefly analyzed.

\section*{Acknowledgements}

This work has been partially supported  by contract $\Phi $-523 with the
National Foundation for Scientific Research, the Ministry of Education,
Science and Technology of Bulgaria.

\end{document}